\documentclass[superscriptaddress,aps,preprint,amsmath,amssymb,floatfix]{revtex4-1}
\usepackage{color}
\usepackage{graphicx}
\usepackage{subfigure}
\usepackage{color}
\usepackage{bm}

\newcommand{\magenta}[1]{\textcolor{magenta}{#1}}

\newcommand{\bS}[0]{{\bf S} }
\newcommand{\bk}[0]{{\bf k} }
\newcommand{\bp}[0]{{\bf p} }
\newcommand{\bK}[0]{{\bf K} }
\newcommand{\bQ}[0]{{\bf P} }
\newcommand{\bX}[0]{{\bf X} }
\newcommand{\btk}[0]{{\bf \tilde{k}} } 
\newcommand{\btq}[0]{{\bf \tilde{p}} }

\newcommand{\up}[0]{ \uparrow  }
\newcommand{\dn}[0]{ \downarrow }

\begin{document}

\begin{center}

{\large\bf Extended Dynamical Mean Field Theory\\
for Correlated Electron Models}
\\[0.5cm]

%


Haoyu Hu$^{1,2}$,
Lei Chen$^1$,
Qimiao Si$^{1}$

{\em $^1$Department of Physics and Astronomy, 
Rice Center for Quantum Materials,
Rice University,
Houston, TX 77005, USA}
\\

{\em $^2$ Donostia International Physics Center, P. Manuel de Lardizabal 4, 20018 Donostia-San Sebastian, Spain}
\\


\end{center}

\vspace{0.5cm}
{\bf
An overarching question in strongly correlated electron systems is how the landscape 
of quantum phases 
emerges from electron correlations.
The method of extended dynamical mean field theory (EDMFT)
has been developed for clean lattice models of the correlated electrons.
For such models, not only onsite Hubbard-like interactions are important,
but so are
intersite interactions.
Importantly,
the EDMFT method treats the interplay between the onsite and intersite interactions dynamically.
It was initially formulated for models of
the two-band Anderson-lattice type with intersite 
interactions, as well as for the
one-band Hubbard type with intersite Heisenberg-like terms that are often called Hubbard-Heisenberg models.
In the case of Kondo lattice models, the EDMFT method 
incorporates a dynamical competition between the local Kondo and 
intersite
Ruderman-Kittel-Kasuya-Yosida (RKKY) interactions.
In these models, the EDMFT-based analyses 
led to the notion of Kondo destruction,
which has 
played a central role in the understanding of quantum critical heavy fermion metals.
In this article, we summarize 
 the EDMFT method, and survey its applications, 
particularly for
Kondo/Anderson lattice models.
We also discuss the prospect for further developing the 
EDMFT method, as well as for applying it to address the correlation
physics in a variety of new settings. Among the latter 
are the orbital-selective Mott 
physics that arises both in iron-based 
superconductors and in frustrated bulk systems 
with topological flat bands.
}

\newpage






\section{Introduction}

A wide variety of bad metals form the bedrock of strongly correlated electron systems. Often, a bad metal is operationally defined in terms of 
the electrical resistivity at room temperature reaches or exceeds the Mott-Ioffe-Regel limit \cite{Pas21.1,Hus04.1}.
This limit is defined in terms of the mean free path $\ell$,
corresponding to $k_F \ell/2\pi \approx 1$ for each Fermi surface, which signifies a proximity to a Mott localized state \cite{Hus04.1}. A complementary criterion is based on
a sizeable reduction of the optical Drude weight from 
its non-interacting-electron expectation 
\cite{Dre02.1,Qaz09.1,Si16.1}.
Heavy fermion metals and metals that lie in the vicinity of a Mott insulator exemplify bad metals.

Quantum many-electron systems are traditionally studied in terms of a perturbative expansion of the interaction~\cite{Nozieres18}.
The assumption is that the noninteracting electrons are the building blocks for low-energy physics.
A standard model is the one-band Hubbard model, which contains an electron band of width $D$ and an onsite interaction
-- the Hubbard interaction -- of strength $U$. 
The Fermi liquid description applies in an
order-by-order perturbative treatment of $U/D$ to infinite orders. Here, coherent quasiparticles appear as a pole
in the single-particle Green's function $G({\bf k},\omega)$.
When $U/D$ 
becomes of order unity,
non-perturbative effects may develop.
The most famous case is the development of a Mott insulator for a half-filled band~\cite{Imada1998}, in which the electron correlations
drive the quasiparticle weight to zero. Concurrently, a gap develops in the single-particle and charge
excitation spectrum; the ground state is no longer a metal and is no longer adiabatically connected to its noninteracting 
counterpart.

In general, when the electron correlations reach and exceed  the electron bandwidth, as is the case in bad metals,
other degrees of freedom
emerge as a part of the building blocks for the low-energy physics. A typical case involves spins,
as seen from the notion of local moment formation
of a correlated and deep orbital in a metallic matrix
\cite{Anderson1961}.
As a result, 
models of strongly correlated electrons may involve both local moments and itinerant electrons, 
as exemplified by heavy fermion metals \cite{Pas21.1,KirchnerRMP,StewartRMP,Coleman-Nature,Si_Science10}.
A similar set of building blocks develop once a 
 Mott insulator~\cite{Imada1998} is coupled to a metallic band.

When expressed in terms of the electron degrees of freedom, the correlation physics describes the effect of Coulomb repulsion
between the electrons \cite{GeorgesRMP,Vollhardt2019}. 
However, if represented in terms of the emergent building blocks such as local moments,
the competition between different types of effective interactions becomes particularly explicit. 
A case in point is a Kondo
lattice Hamiltonian, which contains an antiferromagnetic spin-exchange (Kondo)
interaction between the local moments and 
spins of the conduction electrons 
on the one hand, and the RKKY interaction between the local moments on the other.

The EDMFT method was introduced to dynamically study this type of competition \cite{Si.96,SmithSi-edmft,Chitra2001}.
Consider the case of the Kondo lattice model. Here, 
the local Kondo interaction
favors the formation of a Kondo singlet between the local moments and spins of the conduction electrons.
Whereas the RKKY interaction, which often is also antiferromagnetic,
promotes singlet formation among the local moments.
To properly account for the amplified quantum fluctuations that develop when
the two types of effects would produce comparable energy 
scales, 
the dynamical interplay between the two interactions 
is crucial; this interplay is captured by the EDMFT method. This is to be contrasted with the
standard dynamical mean field theory
\cite{GeorgesRMP,Vollhardt2019}, in which
 the local interactions such as the Kondo couplings are  treated in a dynamical way, 
while the intersite interactions 
such as the RKKY couplings are handled in terms of a static Hartree-Fock approximation.

The EDMFT method has played a central role in the elucidation of quantum phase transitions of heavy fermion metals
\cite{Hu-QCM-review2022}.
Studies based on
such a method have led to the advancement of the concept of Kondo destruction \cite{Si-Nature,Si2003,Si1999,Colemanetal,senthil2004a}.
Qualitatively, the
Kondo physics underlies the formation of (heavy)
quasiparticles in such systems. The quasiparticles
develop through the formation of the Kondo singlet,
and appear in the form of a Kondo-driven composite
fermion. These processes 
are captured by the local self-energy of the 
conduction electrons. 
As such, studying the dynamical
competition between the RKKY and Kondo interactions provides a way of describing 
the reduction -- and eventual destruction -- of the 
Kondo singlet in the ground state and, by extension, the quasiparticles. In other words, the EDMFT approach is ideally suited to access how (heavy) quasiparticles are 
lost.

The concept of Kondo destruction
has influenced the development of quantum critical metals in a profound way, particularly on the destruction
of quasiparticles and dynamical Planckian ($\hbar \omega/ k_{\rm B}T$) scaling at the quantum critical 
point (QCP), and a sudden jump of a ``large'' to ``small" Fermi surface across the QCP \cite{Hu-QCM-review2022}.

The present article is devoted to three topics:

\begin{itemize}

\item
We describe the dynamical equations associated with the EDMFT method.

\item We illustrate the application of the EDMFT approach by focusing on the Kondo lattice models and 
summarizing 
i) the methodological aspect 
of the approach
and ii)
the results as pertaining 
to the heavy fermion quantum criticality. Further analyses that have been motivated by the result of such calculations are touched upon,
especially in the form of a global phase diagram.
\item The prospect for further progress along this general direction is discussed.
\end{itemize}


\section{EDMFT Approach}

We illustrate the EDMFT approach in terms of the
single-band 
Hubbard-Heisenberg-type model
\cite{Si.96,SmithSi-edmft,Chitra2001,ZhuGrempelSi}:

\begin{eqnarray}
H_{\text{U-v}} =
\sum_{i} U n_{i \uparrow} n_{i \downarrow} + 
\sum_{\langle ij\rangle,\sigma} t_{ij} c_{i\sigma}^{\dagger}c_{j\sigma}
+ {1 \over 2} \sum_{\langle ij\rangle } \,v_{ij} \, :n_{i}:
:n_{j}:
\label{eq:hubbard-heisenberg}
\end{eqnarray}
The first two terms specify the Hubbard model
for a spin-$\frac{1}{2}$ band. The onsite Hubbard interaction is  $U$ and the hopping matrix is 
$t_{ij}$, whose Fourier transform corresponds to the band dispersion $\varepsilon_{\bf k}$. The third
term 
describes an intersite
density-density ($v_{ij}$) 
interaction
(a spin-exchange interaction, $J_{ij}$, can also be added, as originally done; see below), with $n_i$ 
being the density operator 
of the $c-$electrons and $:n:\equiv n - \langle n\rangle$ representing its normal-ordered form.
For concreteness, 
we limit the intersite interactions to the nearest-neighbor ($\langle ij \rangle$) contributions, but this can be readily generalized.

\begin{figure}[t!]
\centering
\includegraphics*[width=0.8\textwidth]{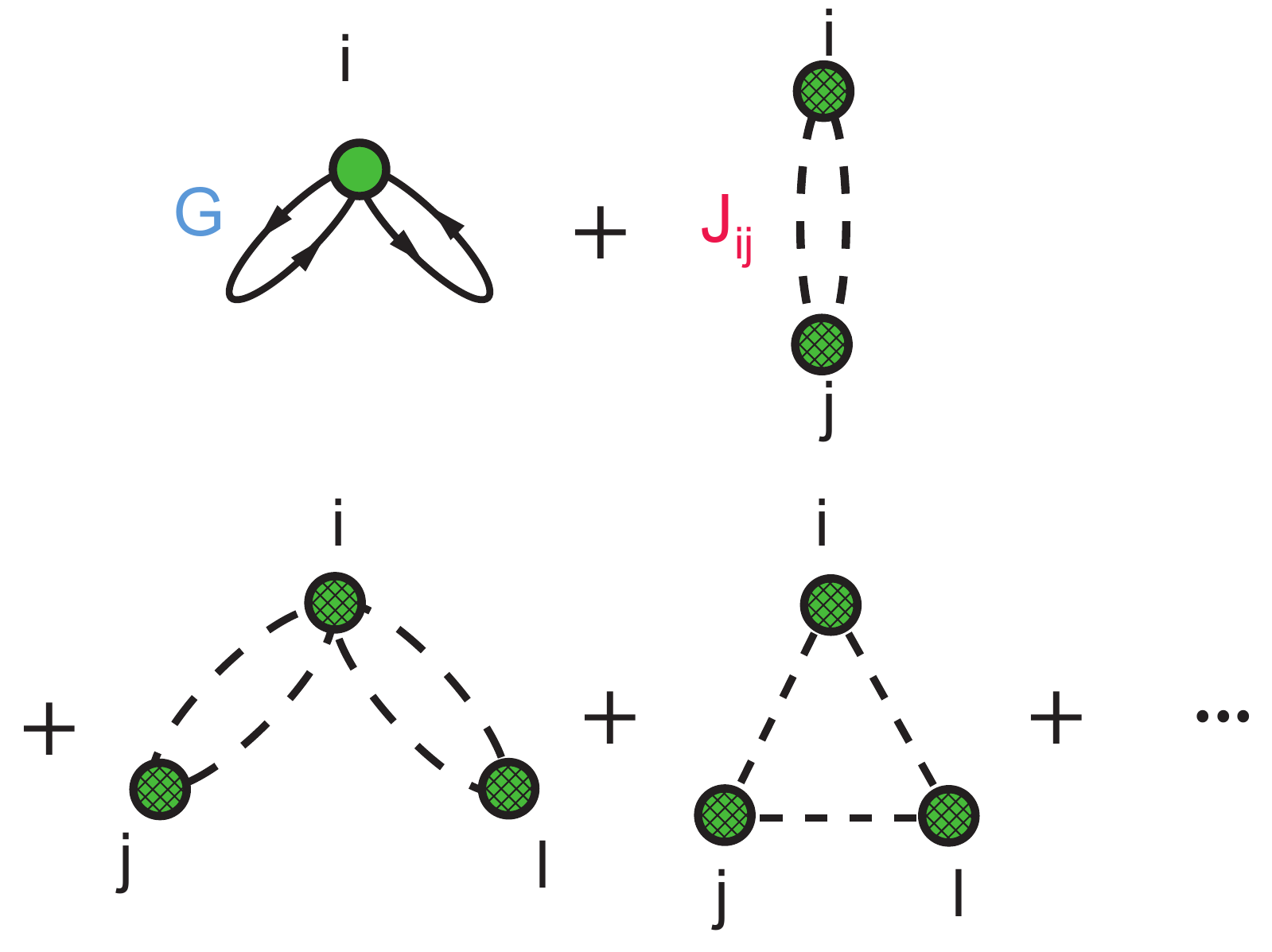}
\caption{\label{fig:diag} 
The single-site, two-site, and three-site diagrams for the Luttinger Ward potential in the extended DMFT~\cite{SmithSi-edmft,Si2003}. A shaded circle contains all the on-site diagrams fully-dressed by the fermion Green's function (solid lines) as shown in the first diagram. The dashed lines further represent the intersite interactions; we have denoted them by $J_{ij}$, but they could also be interactions in other channels, such as $v_{ij}$ of Eq.\,(\ref{eq:hubbard-heisenberg}).}
\end{figure}

The EDMFT approach amounts to the summation of an infinite series of diagrams as outlined in Fig.\,\ref{fig:diag}. The approach is systematic and conserving. Moreover, the EDMFT equations are generated
by an effective action function of the Baym-Kadanoff type.

The EDMFT approach incorporates a local self-energy, $\Sigma$, for the single-electron Green's function $G$ and a related irreducible quantity in the 
density channel, $M$, which is defined in terms of a 
cumulant
that is
$v_{ij}$-irreducible \cite{Si.96};
for notational convenience, it
has been referred to as a 
(density) self-energy. The self-energies determine the dynamical density susceptibility and single-particle Green's function as follows:

\begin{eqnarray}
\chi ({\bf q}, \omega) &=& \frac {1}  { M(\omega) + v_{{\bf q}}} \, ,
\label{eq:chi-n-edmft}
\end{eqnarray}
together with
\begin{eqnarray}
G ({\bf k}, \epsilon) &=& \frac {1}  {\epsilon + \mu
- \epsilon_{\bf k} - \Sigma (\epsilon)} .
\label{eq:g-edmft}
\end{eqnarray}

These self-energies can be calculated from a local action, which can equivalently be expressed in terms of a local Hamiltonian:
\begin{eqnarray}
{\cal H}_{\text{loc,U-v}}
=
U n_{\uparrow} n_{\downarrow}
+ \sum_{\bk,\sigma} E_{p}~c_{\bk\sigma}^{\dagger}~ c_{\bk\sigma}
+ \; g \sum_{\bp} :n:
\left( 
\phi
_{\bp} + 
\phi
_{-\bp}^{\;\dagger} \right)
\label{H-imp}
\end{eqnarray}

Here the dispersion of the bosonic bath, along with its fermionic bath counterpart, are self-consistently determined. The set of nonlinear equations can be expressed as follows. The bath dispersion defines Weiss fields, $\chi_{0}$ and $G_0$, as follows:

\begin{eqnarray}
g^{2} \sum_{\bp}  { 2 w_{\bp} \over {(i\nu_m)^{2} - w_{\bp}^{2}}}
&=& - \chi_{0}^{-1}(i\nu_m),
\nonumber \\[-1ex]
\label{eq:weiss_defs} \\[-1ex]
\sum_{\bk} { 1 \over {i\omega_n - E_\bk}} &=& G_0(i \omega_n) ,
\nonumber
\end{eqnarray}
where $\nu_m$ and $\omega_n$ are
bosonic and fermionic Matsubara frequencies, respectively.
The Dyson equations for the density and electron
self-energies take the following forms:

\begin{eqnarray}
M(\omega)=\chi_{0}^{-1}(\omega) + 1/\chi_{\text{loc}}(\omega) \,, 
\label{eq:dyson-M}
\end{eqnarray}
and, as usual,
\begin{eqnarray}
\Sigma(\omega)=G_0^{-1}(\omega) - 1/G_{\text{loc}}(\omega)
\, .
\label{eq:dyson-sigma}
\end{eqnarray}

The self-consistency equations can then be simply written to capture the translational invariance:
\begin{eqnarray}
\chi_{\text{loc}} (\omega) &=& \sum_{\bf q} \chi ( {\bf q},
\omega ) ,
\nonumber \\[-1ex]
\label{self-consistent} \\[-1ex]
G_{\text{loc}} (\omega) &=& \sum_{\bf k} G( {\bf k}, \omega )\;.
\nonumber
\end{eqnarray}

It can be inferred from Eq.\,\ref{eq:chi-n-edmft} that, for the intersite interactions $v_{ij}$ or $I_{ij}$ (for later reference), 
the following quantity,
\begin{eqnarray}
\label{rho-I} 
\rho_{\rm I} (x) = \sum_{\bf q} \delta (x - I_{\bf q}) \, .
\end{eqnarray}
comes into the self-consistency equation. In practice two particular forms have been used. In one form, $\rho_I(x)$ has a constant onset at the lower edge of its support. In another, it has a square-root onset.

We stress that the dynamical susceptibility is defined in terms of connected correlation functions. As such, the density operator appears in the equations 
in the normal-ordered form. 
This point is also important when states with a 
conventional long-range order (such as charge or spin orders) 
are considered. We will write down the 
corresponding equations below, in the context of 
Anderson/Kondo lattice models. 
The importance of this normal ordering in the EDMFT formulation
was stressed from
the beginning \cite{Si.96,SmithSi-edmft,ZhuGrempelSi}.
When the normal ordering is not taken into account,
the ordered state would yield an action that (unphysically)
scales as $\beta^2$ \cite{ChowdhuryRMP}.


\section{Kondo lattice model: EDMFT equations and solution methods}

The Kondo lattice model represents the limit of the Anderson lattice model when the $f$-electron level lies sufficiently below the Fermi energy, and when the interaction $U$ for the $f$-electrons is sufficiently large \cite{Hewson}. It features two competing interactions. The Kondo coupling/hybridization favors the formation of a Kondo singlet. Whereas the
RKKY interaction leads to the development of quantum magnetism; the typical case corresponds to antiferromagnetic (AF) RKKY interactions, which promotes the formation of inter-moment spin singlets. 

The vantage point of the Kondo effect provides a particularly clear physical picture. It is the development of the Kondo singlet in the ground state that supports heavy quasiparticles in the excitation spectrum. The dynamical competition brought on by the RKKY interactions is detrimental to the quasiparticles. This has led to the notion of Kondo destruction~\cite{Pas21.1,StewartRMP,Coleman-Nature,Si_Science10,KirchnerRMP,Si-Nature}, which has played a central role in the elucidation of quantum critical heavy fermion systems.

In this section, we summarize the basic equations of the Kondo lattice model within the EDMFT approach, and discuss the corresponding solutions for the phases and quantum phase transitions.

\subsection{Kondo lattice Hamiltonian}

The 
Kondo lattice Hamiltonian takes the following form:
\begin{eqnarray}\label{eq:KLHamiltonian}
H_{\rm KL}=
 \sum_{ ij } t_{ij}
c^{\dagger}_{i\sigma}
c^{\phantom\dagger}_{j\sigma}
+
 \sum_{ ij } I_{ij}
{\bf S}_i \cdot {\bf S}_j
+  \sum_{i} J_K {\bf S}_i \cdot c^{\dagger}_{i}
\frac{\vec{\sigma}}{2} c^{\phantom\dagger}_{i} .
\label{eq:kondo-lattice-model-SU2}
\end{eqnarray}
It contains a conduction-electron band, $c_{i\sigma}$,
with hopping matrix $t_{ij}$,
and,
correspondingly, band dispersion
$\varepsilon^{\phantom\dagger}_{\bf k}$.
At each site $i$, the spin of the conduction
electrons, ${\bf s}_{c,i} = (1/2) c_{i}^{\dagger} \vec{\sigma} c_i$,
is coupled to the spin of the
local moment, $\bf {S}_i$,
by a Kondo interaction, namely an AF
Kondo exchange interaction $J_K$.
Here $\vec{\sigma}$ represents the Pauli matrices.
Meanwhile, the local moments are coupled to each other via an RKKY interaction $I_{ij}$.


\subsection{EDMFT equations}

For the Kondo lattice Hamiltonian, the EDMFT approach calculates local correlation function in terms of the following local action
 \cite{Si-Nature,Si2003,ZhuGrempelSi,Glossop07,Zhu07}:

\begin{eqnarray}
{\cal S}_{\text{loc}}
&=& {\cal S}_{\text{top}}
+\int_0^{\beta} d \tau ~J_K ~{\bf S} \cdot {\bf s}_{c} \nonumber\\
&&- \int_{0}^{\beta} d \tau
\int_{0}^{\beta} d \tau' \left[ \sum_{\sigma} c_{\sigma} ^ {\dagger} (\tau)
G_0^{-1}(\tau - \tau ') c_{\sigma}(\tau') \right. \nonumber\\
&& \left. + \; 
{1 \over 2}
:{\bf S}:(\tau) \cdot \chi_{0}^{-1}(\tau - \tau') :{\bf S}:(\tau')
\rule[-2ex]{0ex}{4ex}\right] 
+ \int_{0}^{\beta} d \tau h_{\text{loc}} \, S^z \, .
\label{S-imp-kondo-lattice}
\end{eqnarray}
Here, $\beta = 1/k_B T$, ${\cal S}_{\text{top}}$ describes the
Berry phase
of the local moment, while $G_0^{-1} $ and $\chi_{0}^{-1}$ are the 
dynamical Weiss fields determined self-consistently by Eq.~\ref{self-consistent}. Finally, $h_{loc}$ is the static Weiss field determined by an additional self-consistent equation:
\begin{eqnarray} 
h_{loc} = -I_{\textbf{Q}} \langle S^z\rangle \, . 
\end{eqnarray} 
The retarded interactions of the spin fields have been written with normal ordering.
Equivalently, it could be expressed without using normal ordering:
\begin{eqnarray} 
&&-\int_0^\beta d\tau \int_0^\beta d\tau' \frac{1}{2}:{\bf S}:(\tau) \cdot \chi_{0}^{-1}(\tau - \tau') :{\bf S}:(\tau')\nonumber \\
&=& -\frac{1}{2}{\bf S}(\tau) \cdot \chi_{0}^{-1}(\tau - \tau'){\bf S}(\tau') +\int_0^\beta d\tau \tilde{h}_{loc}  S^z(\tau) 
+ const
\end{eqnarray} 
$\tilde{h}_{loc} =\int_0^\beta \chi_0^{-1}(\tau) m^z d\tau $ is the effective magnetic field from normal ordering and $m^z = \langle S^z\rangle$.
This action is equivalent to the following effective local 
Hamiltonian:
\begin{eqnarray}
H_{BFK} &= &\sum_{ \bk }E_{\bk}
c^{\dagger}_{\bk\sigma}
c^{\phantom\dagger}_{\bk\sigma} 
+ \sum_{\bp} \omega_{\bp} \bm{\phi}_{-\bp}^{\dag}
\bm{\phi}_{\bp}
\nonumber 
\\
&& + J_K \bm{S} \cdot \frac{c_0^\dag \bm{\sigma} c_0}{2} + g :\bm{S}: \cdot \sum_{\bp}( \bm{\phi}_{\bp}^{\dag}+ \bm{\phi}_{-\bp}^{\dag}) 
+ h_{loc} S^z 
\end{eqnarray}
where $E_{\bk}$ and $\omega_\bp$ characterize the 
dispersion of the fermionic fields ($c_{\bk,\sigma}$) and bosonic fields ($\bm{\phi}_{\bp}
$), and $J_K$ ($g$) is the coupling strength between the impurity and fermionic fields (bosonic fields). The relation between the 
dispersions and Weiss fields are given in Eq.~\ref{eq:weiss_defs}.

\subsection{Variants of the Kondo lattice model}

The Kondo lattice Hamiltonian, Eq. \ref{eq:kondo-lattice-model-SU2}, has an SU(2) symmetry.
In heavy-fermion systems, 
the spin-orbit coupling can reduce the spin symmetry down to either easy plane (xy-anisotropic) or easy axis (Ising-anisotropic). 
This can be treated by making the interactions to depend on the spin component $\alpha=x,y,z$.
Because the flow towards the Kondo fixed point in systems with anisotropic Kondo interactions restores the SU(2) symmetry, in practice it suffices to incorporate the spin-anisotropy in the RKKY interaction, namely 
to replace 
the term $ \sum_{ ij, \alpha=x,y,z} I_{ij}
{\bf S}_i \cdot {\bf S}_j $ by
$\sum_{ ij } I_{ij}^{\alpha}
S^{\alpha}_i \, S^{\alpha}_j $. Correspondingly, in
the Bose-Fermi Kondo model, both the dispersion of the
bosonic bath and the Bose-Kondo coupling $g$ depends on the component $\alpha=x,y,z$.

For large but finite values of the Hubbard interaction $U$ and $f$-orbital-energy level separation from the Fermi level, the $f$ electrons behave as correlated electrons instead of a pure quantum spin. The system in this case is described by an Anderson lattice model with explicit RKKY interactions:
\begin{eqnarray}
H_{\rm AL} = 
 \sum_{ ij } t_{ij}
c^{\dagger}_{i\sigma}
c^{\phantom\dagger}_{j\sigma}
+
 \sum_{ ij } I_{ij}
{\bf S}_i \cdot {\bf S}_j 
+  \sum_{i \sigma} V (f_{i\sigma}^\dag c_{i\sigma} + c_{i\sigma}^\dag f_{i\sigma})  + \sum_i (U n_{i\up}^f n_{i\dn}^f + \epsilon_d n_i^f ) \, .
\label{eq:AI}
\end{eqnarray}
Here, $f_\sigma^\dag$ creates a correlated electron with spin $\sigma$, $n_{i\sigma}^f = f_{i\sigma}^\dag f_{i\sigma}$ and ${\bf S}_i =\frac{f^\dag_i \bm{\sigma} f_i}{2}$ are the density operator and spin operators of the $f$ electrons,
$\epsilon_d$ is the local energy scale, and $V$ is the hybridization strength. Via EDMFT, the local 
correlators of the Anderson lattice model are
determined by a Bose-Fermi Anderson model 
with the following action
\begin{eqnarray} 
S_{\text{loc,A}} = 
&&- \int_{0}^{\beta} d \tau
\int_{0}^{\beta} d \tau'  \sum_{\sigma} 
\left[
c_{\sigma} ^ {\dagger} (\tau)
G_0^{-1}(\tau - \tau ') c_{\sigma}(\tau') 
+f_\sigma^\dagger (\tau) (\partial_\tau-\epsilon_d) f_\sigma(\tau) 
\right] \nonumber\\
&&+ \int_0^\beta d\tau \left[
Un_\up^f (\tau)n_\dn^f(\tau) + V\sum_\sigma( c_\sigma^\dagger(\tau)f_\sigma(\tau) +\text{h.c.})+h_{\text{loc}} \, S^z(\tau) \right] 
\nonumber 
\\ 
&& -\frac{1}{2} \int_0^\beta d\tau \int_0^\beta d\tau' 
:{\bf S}:(\tau) \cdot \chi_{0}^{-1}(\tau - \tau') :{\bf S}:(\tau')
\end{eqnarray}
The corresponding Hamitltonian reads 
\begin{eqnarray}
H_{BFA} &= &\sum_{ \bk }E_{\bk}
c^{\dagger}_{\bk\sigma}
c^{\phantom\dagger}_{\bk\sigma} 
+ \sum_{\bp} \omega_{\bp} \bm{\phi}_{-\bp}^{\dag}
\bm{\phi}_{\bp}
+ h_{loc} S^z +Un_\up^f n_\dn^f + \epsilon_d (n_\up^f+n_\dn^f)
\nonumber 
\\
&& + V\sum_{\bk,\sigma}(c_{\bk \sigma}^\dag f_\sigma +\text{h.c.}) + g :\bm{S}: \cdot \sum_{\bp}( \bm{\phi}_{\bp}^{\dag}+ \bm{\phi}_{-\bp}^{\dag}) 
\end{eqnarray}
where the relation between bath dispersions and bath functions are given in Eq.~\ref{eq:weiss_defs}.

\subsection{Bose-Fermi Kondo model}

\begin{figure}[h!]
\centering
\includegraphics*[width=0.75\textwidth]{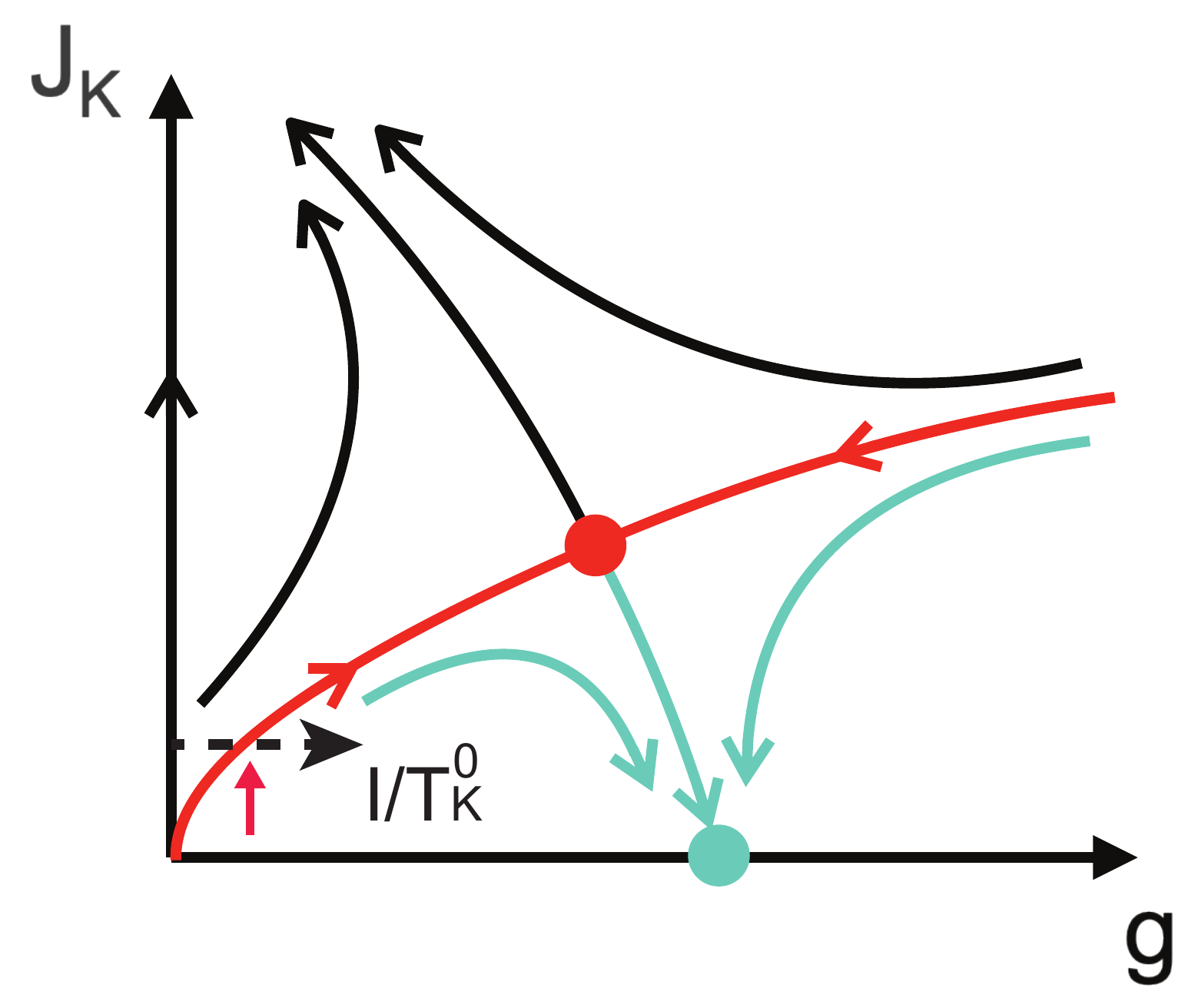}
\caption{\label{fig:RGflow}
The Kondo fixed point and its destruction of the 
Bose-Fermi Kondo model,
described in terms of an
RG flow
based on an expansion in $\epsilon$~\cite{Si2003} (that 
is 
taken to be $> 0$).
}
\end{figure}

The Bose-Fermi Kondo model contains a Kondo fixed point and, for a suitable form of the bosonic-bath spectrum, its destruciton. This was first seen in an RG analysis based on an $\epsilon-$expansion in the Ising-anisotropic Bose-Fermi Kondo model \cite{Si.96}, which was subsequently generalized to the SU(2)-symmetric version of the model \cite{SmithSi_EPL1999,Sengupta}. Here, $\epsilon$ is defined in terms of the power-law spectrum of the bosonic bath:
\begin{eqnarray}
\rho_b(\omega) \equiv \sum_{\bp}  \delta (\omega - w _\bp)  
\propto 
|\omega|^{1-\epsilon} ~~~~~~~~\text{for} ~ |\omega| < \Lambda \, .
\label{dos-boson}
\end{eqnarray}
When $\epsilon=0$, the bosnic bath has an ohmic spectrum. For $\epsilon>0$, the spectrum is sub-ohmic. The RG analysis was 
extended to two and higher loops \cite{ZhuSi,ZarandDemler}.

To one loop, the RG equations are 
\begin{eqnarray}
\beta(J_K) = J_K(J_K-g^2) \quad,\quad 
\beta(g) = g(\epsilon/2- g^2) \, .
\label{eq:rg_eq}
\end{eqnarray}
They yield a Kondo destruction (KD) quantum critical point as shown in Fig.~\ref{fig:RGflow}. The KD quantum critical point describes a quantum phase transition between a Kondo-screened phase with $J_K\ne 0$ and a Kondo-destroyed phase with $J_K=0$. 

The Kondo fixed point features a spin-singlet ground state. An infrared Kondo energy scale is nonzero, which makes the local spin susceptibility to take a Pauli form. As the system approaches the separatrix of the RG flow, which is controlled by the critical KD fixed point, this infrared Kondo energy scale collapses to zero, and the local spin susceptibility diverges in the zero-frequency limit at $T=0$. On the other side of the separatrix, the RG flow is towards a fixed point with zero Kondo coupling. Nonetheless, dynamical Kondo effect still takes place here when the bare Kondo coupling is nonzero.

In the SU(2) BFKM, a very recent study has 
analytically treated the model 
based on a large-$S$ expansion~\cite{Hu22.fix}, where $S$ is the spin size. This analysis finds a slew of new fixed points for small $\epsilon$, beyond the ones that were identified in the aforementioned $\epsilon$-expansion RG method. With increasing $\epsilon$, the analysis identifies a sequence of fixed-point annihilation, which are consistent with -- and extends -- the results of quantum Monte Carlo calculations~\cite{Cai-CTQMC19.1}.

\subsection{Solution methods}

To analyze the self-consistent equations of the EDMFT approach, one needs to solve the Bose-Fermi Kondo/Anderson models. We have already discussed the $\epsilon$-expansion RG method. This analytical approach is accompanied by several numerical methods.

One is the numerical renormalization group (NRG) method
\cite{Glossop.05,Glossop07}.
Within the framework of NRG, we first preform a logarithmic discretization of both the fermionic bath and bosoinc bath and map both baths into semi-infinite chains where the local degrees of freedom are coupled to the first site. Subsequently, the model is solved by diagonalizing the Hamiltonian iteratively. In each step, we add new sites to both chains.

Another is the continuous-time Monte Carlo method.
Both the SU(2)-symmetric Anderson model \cite{Cai-CTQMC19.1}  and the Anderson model in the Ising limit \cite{pixley2011,pixley2013}.
can be solved via the quantum Monte Carlo method. For the model in the Ising limit, we first perform the unitary (Firsov-Lang) 
transformation \cite{pixley2011,pixley2013,Werner-2010}
$f\rightarrow \tilde{f}_\sigma = f_\sigma e^{-g/2 \sigma \Phi^z}$, where $\Phi^z$ is a linear function of the bosonic fields corresponding to the bosonic bath. Then we can solve the model by directly expanding the hybridization term $ V (c_\sigma^\dag \tilde{f}_\sigma + \tilde{f}^\dag_\sigma c_\sigma)$ and sampling the configuration via the Monte Carlo method. As for the SU(2)-symmetric Anderson model, besides expanding the hybridization term, we also need to expand the perpendicular coupling
\cite{Cai-CTQMC19.1,Otsuki-2013} between spin impurity and bosonic fields $g S^{x/y} \phi^{x/y}$. 


\section{Kondo lattice model: Kondo destruction}

\subsection{Phase diagram}

\begin{figure}[h!]
\centerline{\hspace{-0.5cm}\includegraphics[height=0.366\textwidth]{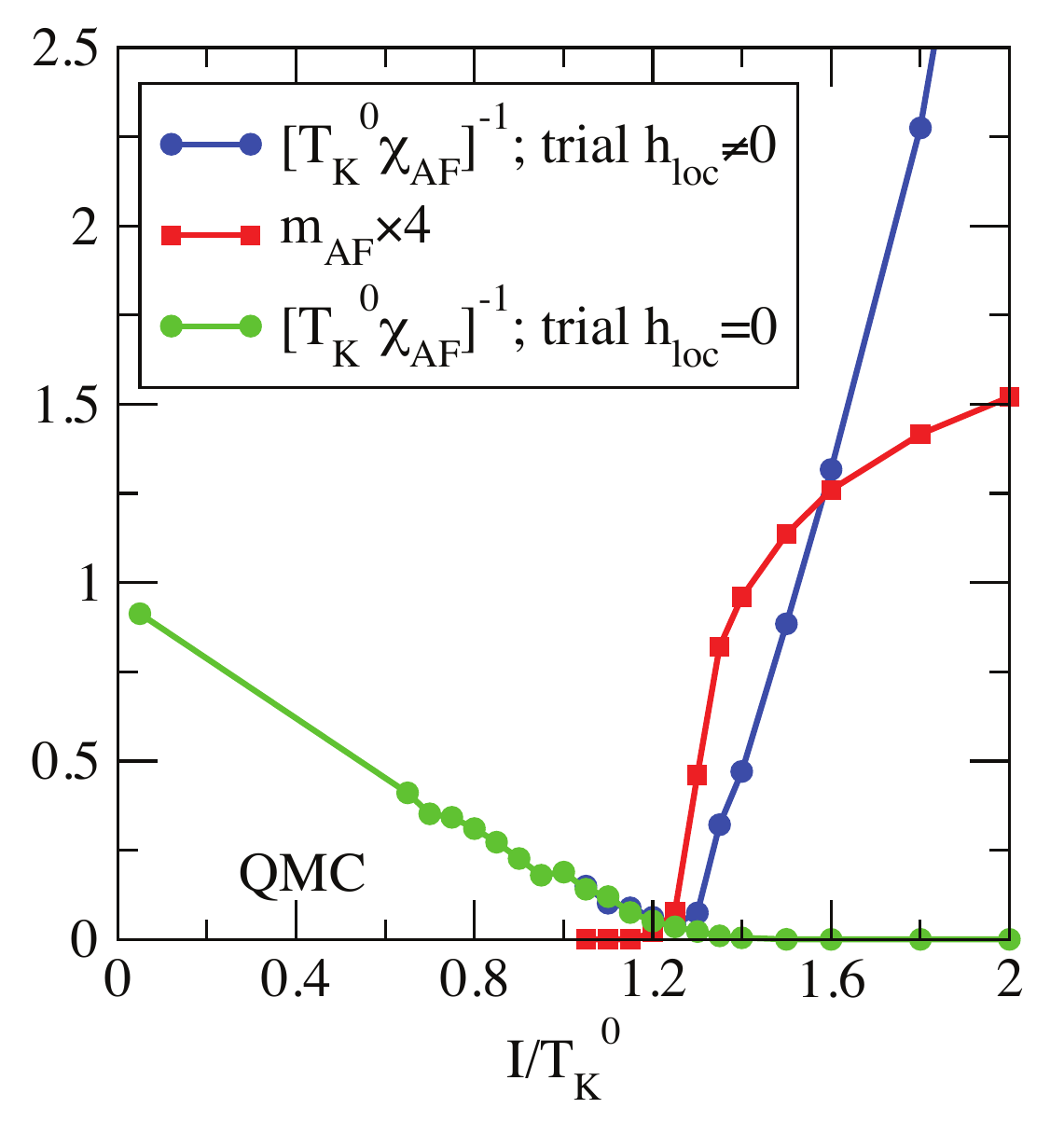}\hspace{0.7cm}\includegraphics[height=0.366\textwidth]{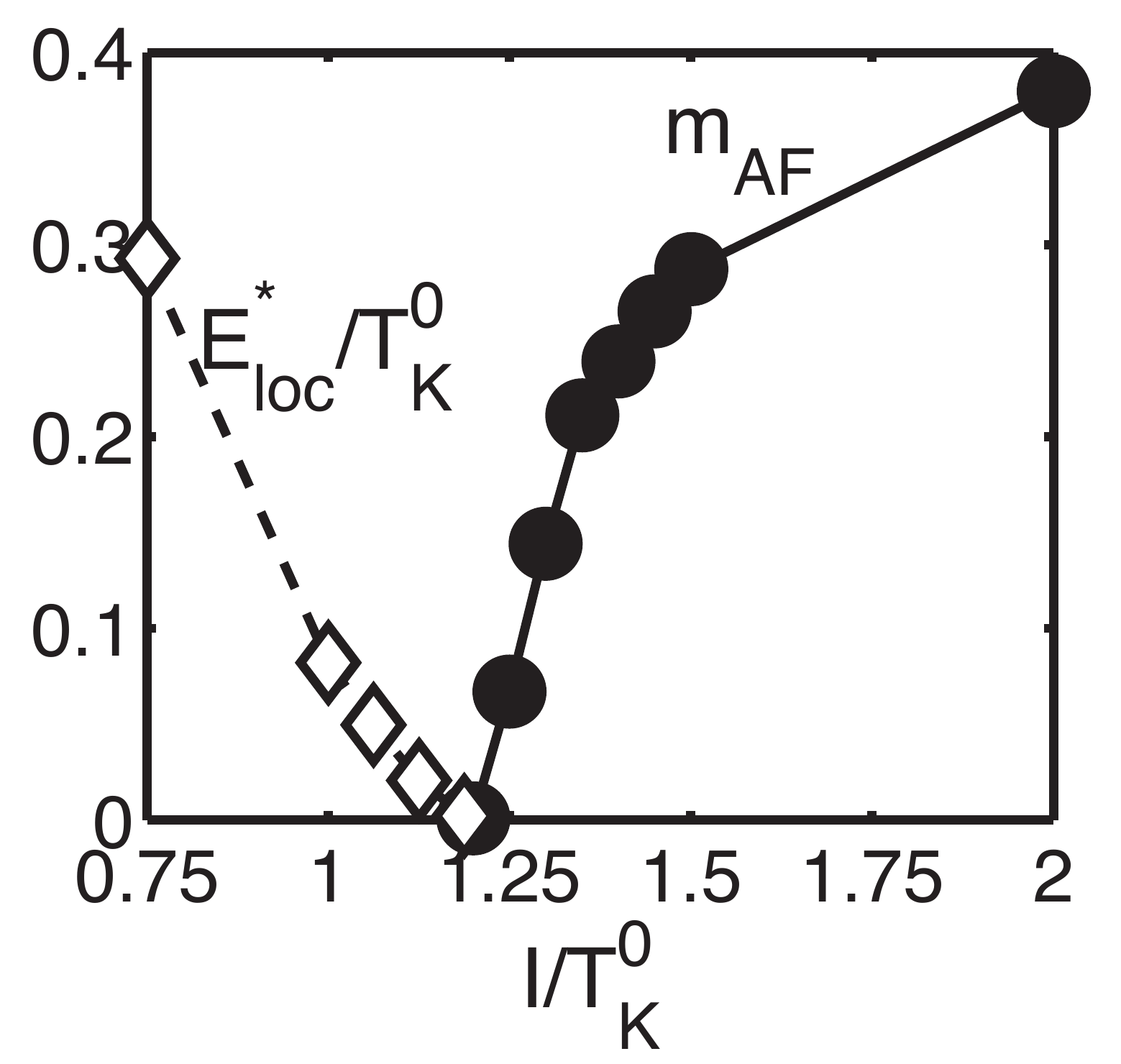}}
\vspace{-6.8cm}

\hspace{-7.2cm}{\bf\textsf{a}}\hspace{6.5cm}{\bf\textsf{b}}
\vspace{6cm}

\caption{\label{fig:Phase}
(a) The phase diagram for the Ising-anisotropic Kondo lattice model as determined by the EDMFT approach.
The inverse static AF susceptibility as a function of $I/T_{K}^{0}$ from the paramagnetic (trial $h_{loc}=0$) and AF (trial $h_{loc}\neq0$) solutions~\cite{Zhu2007}. (b) The staggered magnetization (solid circle) and the Kondo-destruction energy scale $E_{loc}^{*}$ (open diamonds) as a function of $I/T_{K}^{0}$~\cite{Zhu2003PRL}.  }
\end{figure}

The EDMFT analysis led to a continuous zero-temperature transition between a Kondo-driven paramagnetic phase and 
an AF order, as illustrated in Fig.\,\ref{fig:Phase}(a), for an Ising-anisotropic Kondo lattice \cite{Zhu07}.
A $E_{\text{loc}}^{\ast}$ scale is defined in terms of the local responses of the effective Bose-Fermi Kondo model. Qualitatively, a vanishing $E_{\text{loc}}^{\ast}$ means that the Bose-Fermi Kondo model has reached the critical Kondo-destruction fixed point -- the red fixed point shown in the RG flow of 
Fig.\,\ref{fig:RGflow}. 

\subsection{Kondo destruction}

For the 
Ising-anisotropic Kondo lattice model, the scale 
$E_{\text{loc}}^{\ast}$ as a function of the tuning parameter $\delta$, defined as the ratio of the RKKY interaction ($I$) to the bare Kondo scale ($T_{\text{K}}^0$) is shown in Fig.\,\ref{fig:Phase}(b)
\cite{GrempelSi,ZhuGrempelSi}.
The scale $E_{\text{loc}}^{\ast}$ is found to vanish at the QCP, signifying that the QCP is of the Kondo-destruction type. This feature has recently been shown to develop in the SU(2) Kondo lattice model \cite{Hu20.1x}.

\begin{figure}[h!]
\centerline{\hspace{-0.5cm}\includegraphics[height=0.36\textwidth]{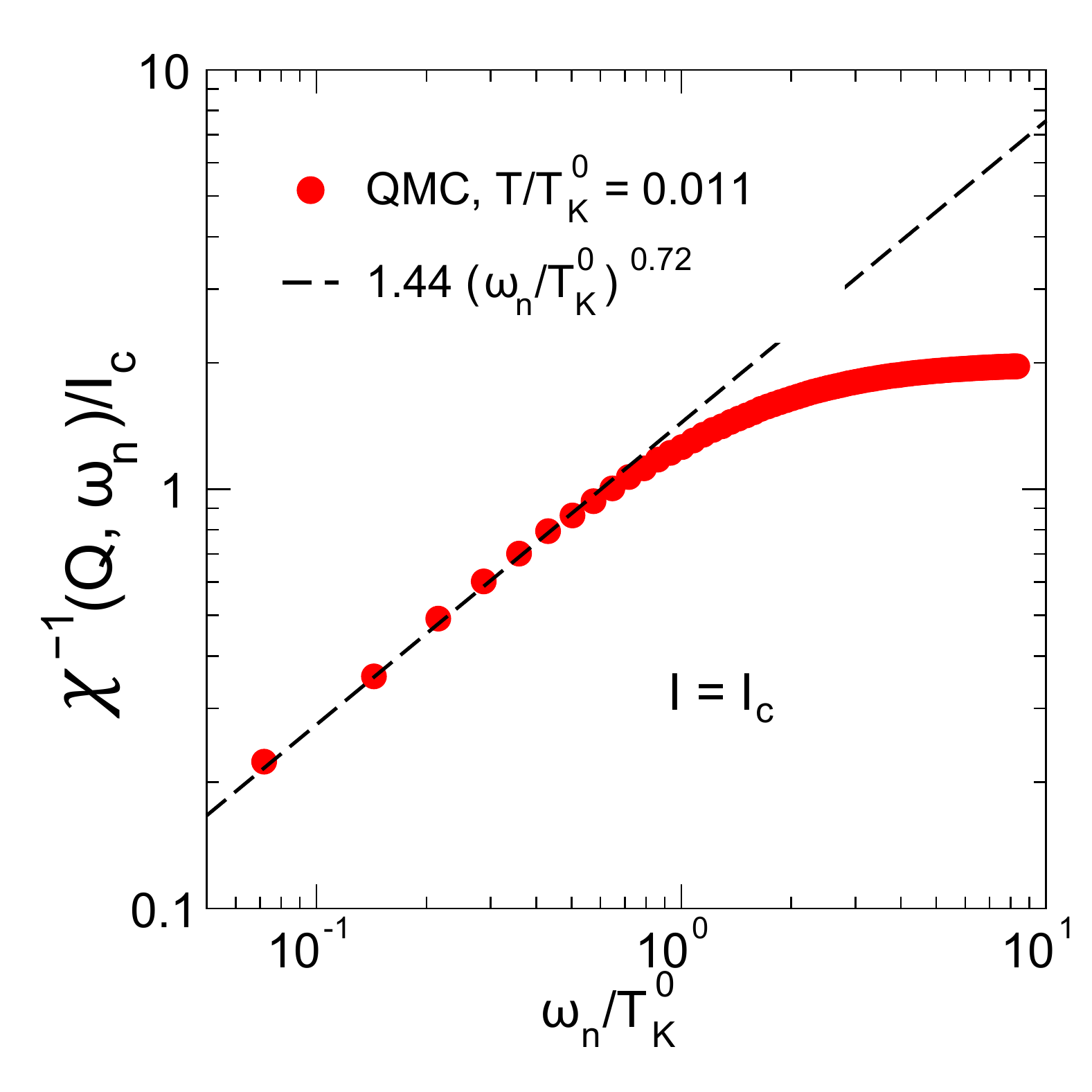}\hspace{0.7cm}\includegraphics[height=0.366\textwidth]{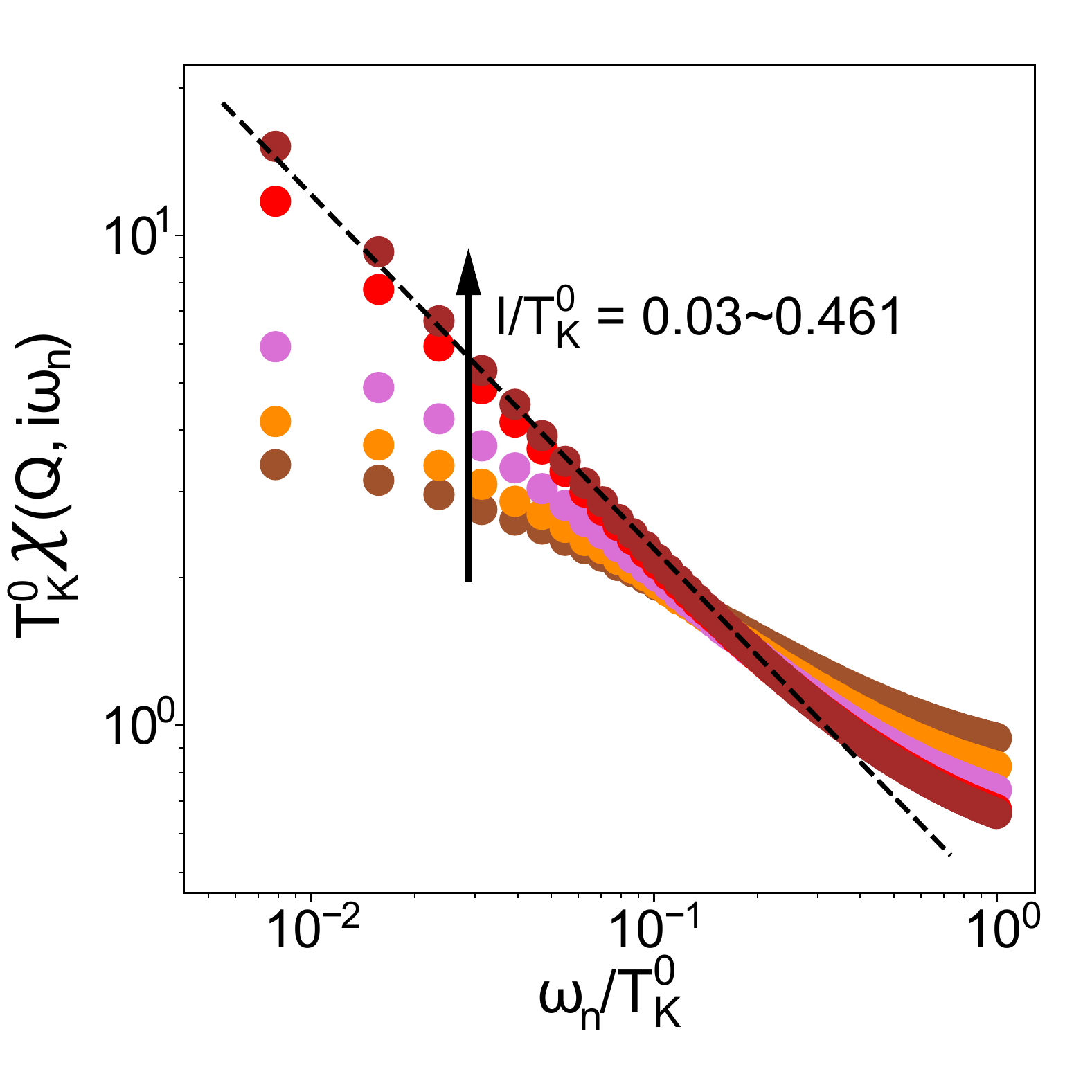}}

\vspace{-6.8cm}

\hspace{-7.2cm}{\bf\textsf{a}}\hspace{6.5cm}{\bf\textsf{b}}
\vspace{6cm}

\caption{\label{fig:SpinRes} Frequency dependence of the antiferromagnetic lattice spin susceptibility for (a) Ising-anistropic Kondo lattice at the QCP~\cite{SiGrempelZhu} and (b) SU(2) Anderson lattice in the paramagnetic side~\cite{Hu20.1x}. }
\end{figure}

Kondo destruction leads to the development of singular
spin dynamics with a fractional exponent at the QCP. 
This is illustrated in Fig.\,\ref{fig:SpinRes}(a) for the Ising-anisotropic Kondo lattice model~\cite{SiGrempelZhu} and in
Fig.\,\ref{fig:SpinRes}(b) 
for the SU(2)-symmetry Kondo lattice model
\cite{Hu20.1x}.

\begin{figure}[h!]
\centering
\includegraphics*[width=0.7\textwidth]{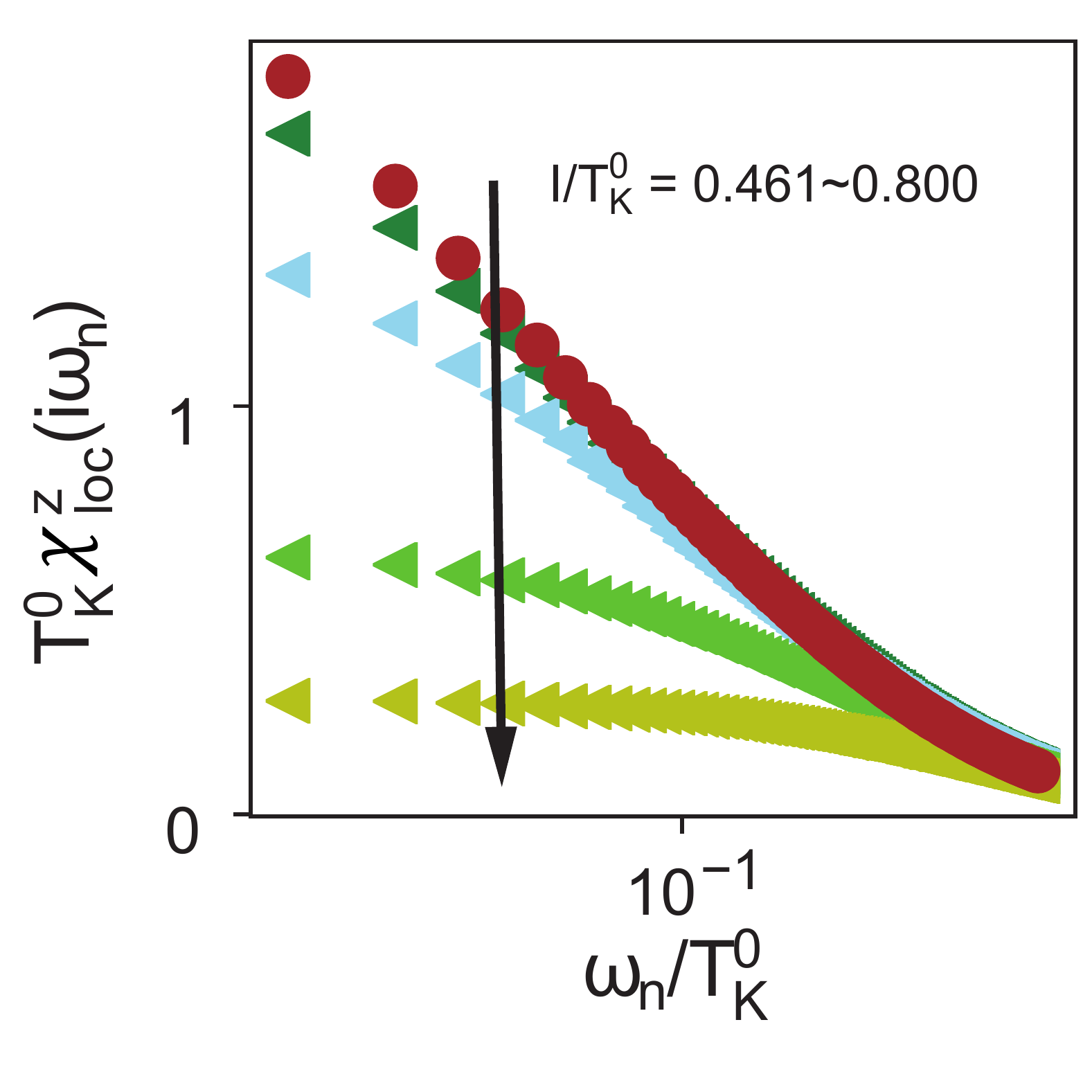}
\caption{\label{fig:LocalChi}
Dynamical local spin susceptibility for SU(2) Anderson lattice model in the magnetic ordered side~\cite{Hu20.1x}. 
}
\end{figure}

\subsection{Dynamical Kondo effect}

Kondo destruction in Kondo lattice models originates from the dynamical competition between the Kondo and RKKY exchange interactions. The exact marginality of the Kondo coupling inside the AF order phase has been demonstrated based on an RG analysis within a quantum nonlinear sigma model analysis
\cite{Yamamoto07,Yamamoto10}.
Indeed, the dynamical Kondo effect allows the gain of a Kondo exchange energy in the absence of static Kondo singlet formation, which is important to help stabilize a Kondo-destroyed phase and allows for the quantum phase transition from Kondo phase to be second order. 
The dynamical Kondo effects 
is particularly pronounced in the Kondo destroyed phase,
in which the amplitude of the {\it static} Kondo-singlet vanishes. 
Figure.~\ref{fig:LocalChi} shows the local dynamical spin susceptibility, $\chi_{loc}(w_n)$, as a function of frequency in the AF phase~\cite{Hu20.1x}. The monotonic increase of $\chi_{loc}(w_n)$ as $I/T_{K}^{0}$ is tuned 
towards the QCP reflects the growth of the dynamical Kondo effect. The match of the local dynamics from both sides at the quantum critical point demonstrates the second-order nature of the zero-temperature transition. The dynamical Kondo effect also implies that the quasiparticles near the small Fermi surface in the Kondo-destroyed phase have an enhanced mass, which is important to understand the enhancement of the Sommerfeld coefficient, cyclotron mass, and quadratic-in-$T$ coefficient of the electrical resistivity in the Kondo-destroyed phase implicated in YbRh$_2$Si$_2$
(Ref.\,\cite{Gegenwart02}) and related systems.


\section{Kondo lattice Hamiltonian: Singular charge response}

\begin{figure}[h!]
\centerline{\hspace{-0.5cm}\includegraphics[height=0.366\textwidth]{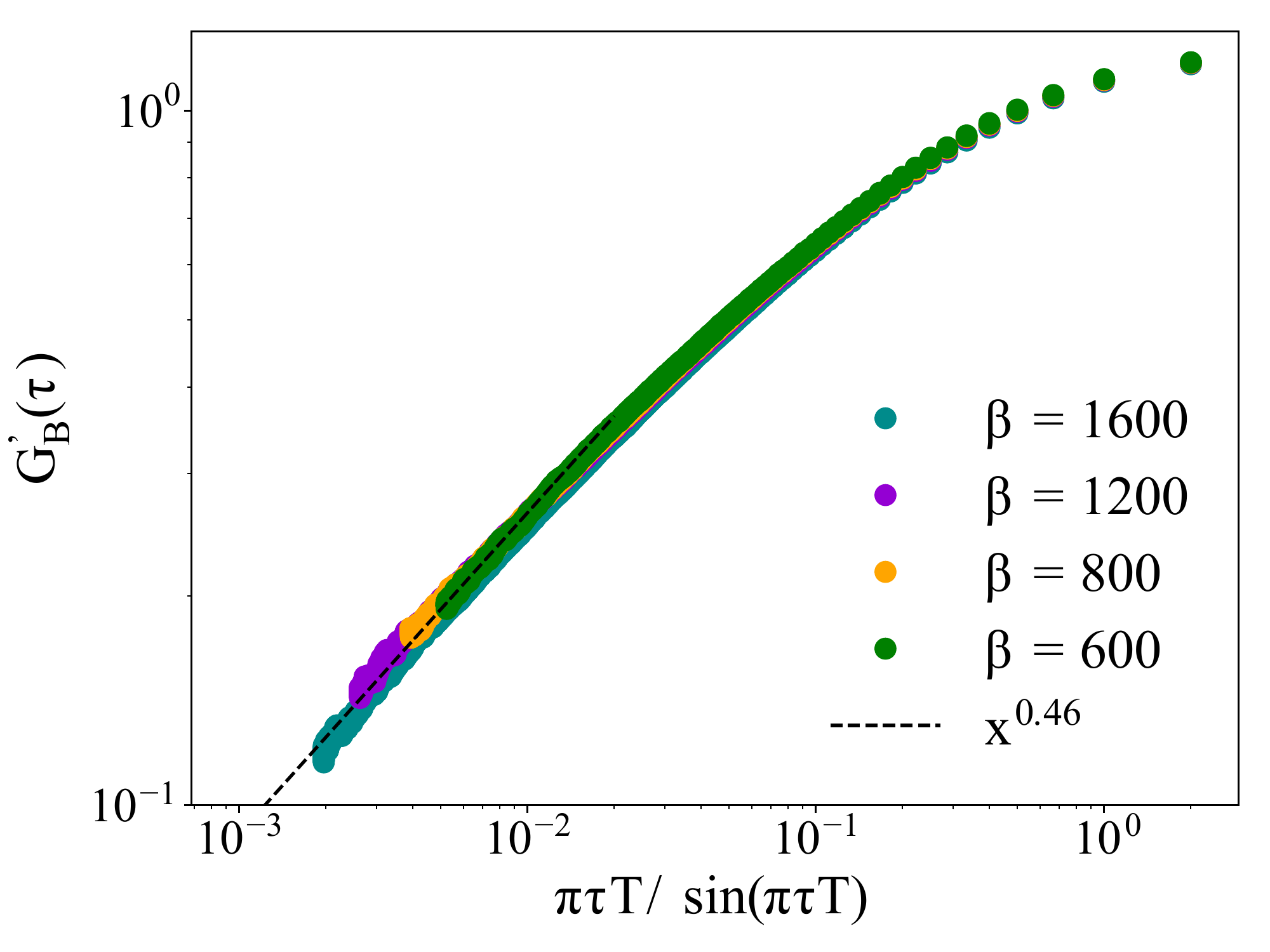}\hspace{0.7cm}\includegraphics[height=0.366\textwidth]{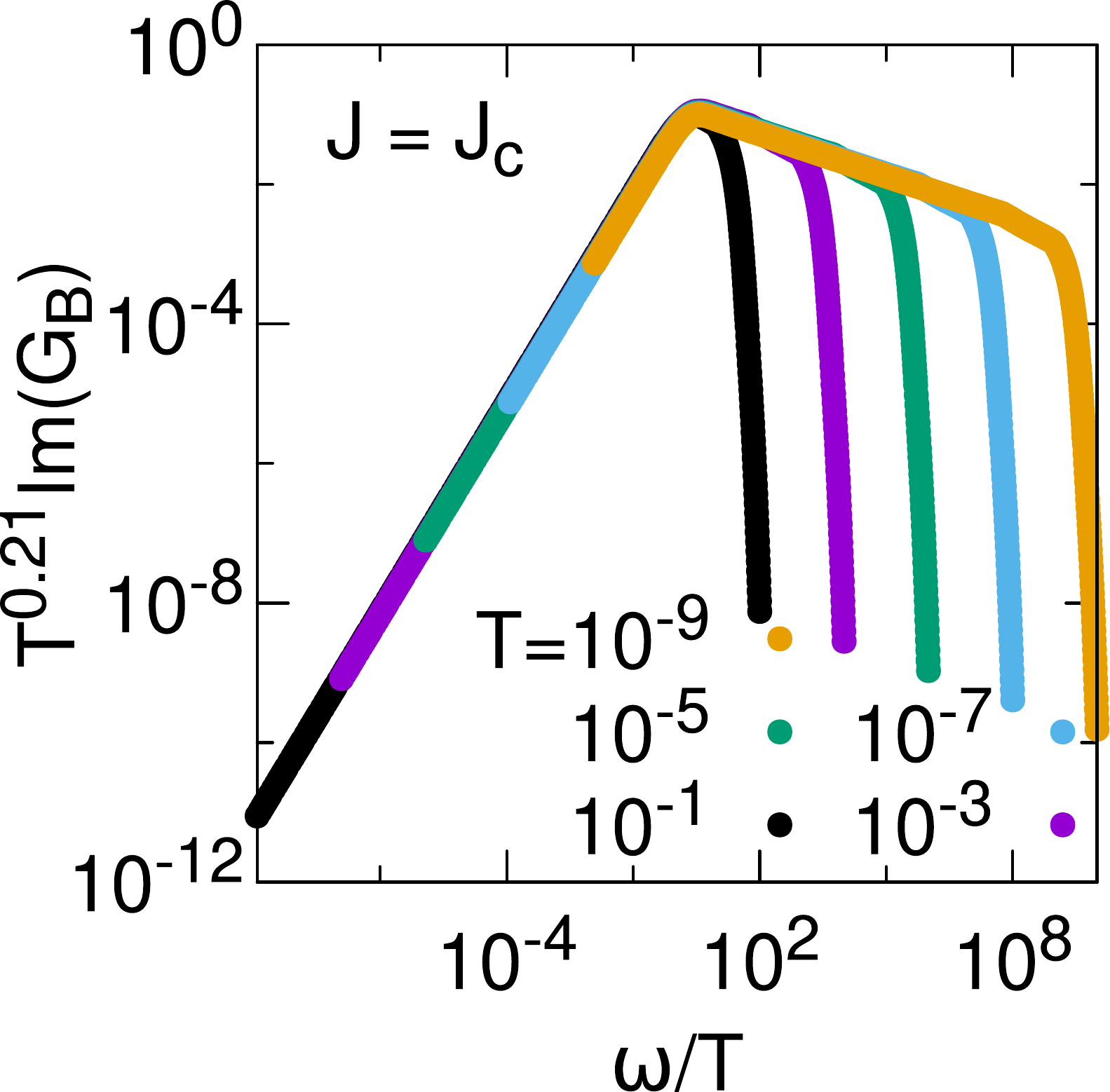}}
\vspace{-6.8cm}

\hspace{-6.2cm}{\bf\textsf{a}}\hspace{8.5cm}{\bf\textsf{b}}
\vspace{6cm}

\caption{\label{fig:scaling} (a) Singular charge response of the hybridization $B$-field in the lattice model at the KD quantum critical point, calculated by the EDMFT method. (b) Dynamical Planckian scaling of the charge response at the KD quantum critical point 
in the spectral function of the $B$-field in the dynamical large-$N$ limit. Both figures from Ref.\,\cite{Ang19PRL} (see also Ref.\,\cite{Zhu.04}). }
\end{figure}

In addition to the spin dynamics, Kondo destruction quantum criticality also involves the localization-delocalization of the $f$-electrons at the QCP. This partial Mott transition is accompanied by a sharp large-to-small Fermi surface reconstruction across the phase transition. Thus, the charge degree of freedom represents an integral part of the quantum criticality. A recent terahertz spectroscopy measurement in
the quantum critical heavy fermion metal 
YbRh$_2$Si$_2$ have discovered a singular charge response  \cite{Prochaska2020}. This observation is inconsistent with conventional SDW QCP, where only the order parameter should be singular and the charge correlations are expected to be smooth. The EDMFT analysis 
is able to capture the singular charge response at the KD QCP~\cite{Ang19PRL}.  As shown in Fig.~\ref{fig:scaling}(a), the charge-carrying hybridization B-field at the KD QCP 
follows dynamical Planckian scaling. This result is 
complemented by the corresponding calculation in a Bose-Fermi Kondo model within 
a dynamical large-$N$ approach, where the index $N$ 
appears in the spin channel, 
as shown in Fig.~\ref{fig:scaling}(b).

\section{Further developments}

We now turn to discussing the prospect for further developing the 
EDMFT method, as well as for applying it to new settings for correlation
physics.

\subsection{Cluster version of EDMFT}
\begin{figure}[h!]
\includegraphics*[width=0.8\textwidth]{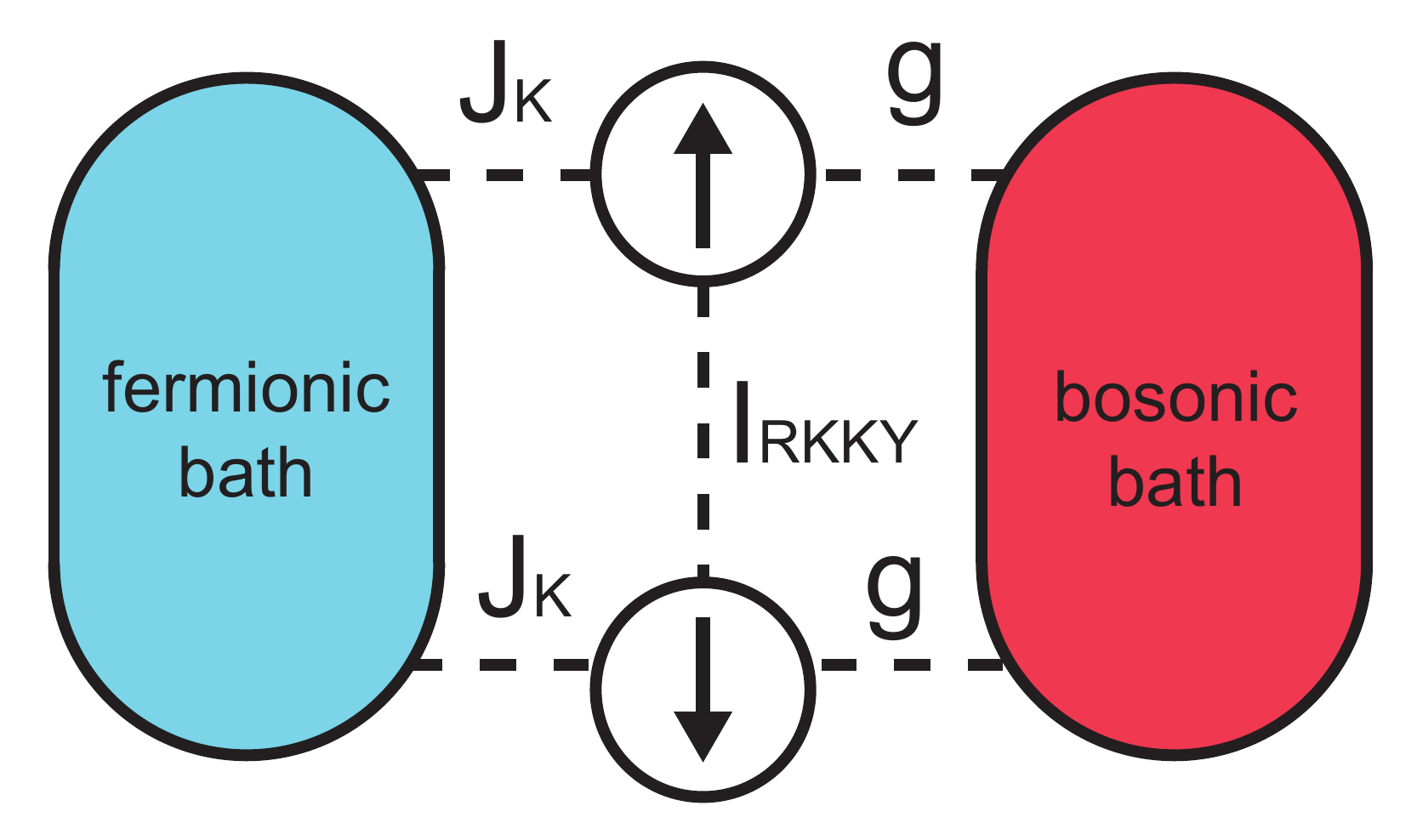}
\caption{\label{fig:cluster} Illustration of the two impurity Bose -Fermi Anderson model that arises from the CEDMFT approach. }
\end{figure}

We first summarize the cluster extended dynamical mean field theory (CEDMFT) \cite{CEDMFT15}. We first divide a $d$ dimensional lattice with size $L$ into several clusters, where each cluster has $N_c (=(L_c)^d)$ sites with size $L_c$. We therefore use ${\bf \tilde{{x}}}$ to label the position of cluster and use $\bX$ to label the position of site within the cluster. 
Correspondingly, we let $\bK$ 
(or $\bQ$)
and $\btk$ (or $\btq$)
to label the reciprocal spaces corresponding to the position vectors $\bX$ and ${\bf \tilde{{x}}}$ respectively. Therefore, the original momentum vector $\bk$ can be written as $\bk = \bK + \btk$.
Then, via dynamical cluster approximation, we are able to map the original lattice model to a self-consistent impurity model, where the impurities are defined on a cluster. In the case of Kondo lattice model defined in Eq.~\ref{eq:KLHamiltonian}, we obtain the following impurity model
\begin{eqnarray} 
S_{imp} &=& S_c + S_\phi +S_I \nonumber  \\
S_c&=& - \int_0^\beta \int_0^\beta \sum_{\bK, \sigma} c_{\bK,\sigma}^\dag(\tau) G^{-1}_{0}(\tau-\tau',\bK) c_{\bK,\sigma}(\tau')d\tau d\tau' \nonumber  \\
S_{\phi}&=& -\frac{1}{2} \int_0^\beta \int_0^\beta \sum_{\bQ, \mu} S^\mu_{\bQ}(\tau) \chi^{\mu, -1}_{0}(\tau-\tau',\bQ) S^\mu_{-\bQ}(\tau')d\tau d\tau' \nonumber  \\
S_I &=& \int_0^\beta \sum_{\bQ,\bK}
\frac{J_K}{\sqrt{N_c}} \bS_{\bQ}(\tau) \cdot \frac{ c_{\bK+\bQ}^\dag (\tau) \bm{\sigma} c_{\bK}(\tau) }{2}d\tau 
d\tau + \int_0^\beta \sum_{\bQ}\bar{I}_{\bQ} \bS_{\bQ}(\tau) \bS_{-\bQ}(\tau)  d\tau  \nonumber \\
&& + \int_0^\beta \sum_{\bQ,\mu} h^\mu_{\bQ}  \bS^\mu_{\bQ}(\tau) d\tau 
\label{eq:KLimp}
\end{eqnarray} 
where the intra-cluster interactions are defined as $\bar{I}_{\bQ} = \frac{(L_c)^d}{L^d} \sum_{\btq} I_{\bQ + \btq}$ with $I_{\bp}$ the Fourier transformation of RKKY interactions $I_{ij}$.
$h_{\bQ}^\mu$ corresponds to an effective magnetic field, which characterize the development of long-range magnetic order. $G_0(\tau,\bK)$ and $\chi_0^\mu(\tau,\bQ)$ are the fermionic and bosonic bath functions respectively. The case with $N_c=2$ is illustrated in Fig.\,\ref{fig:cluster}.

Via the Dyson equations, the self-energies of the single-particle Green's function and spin-spin correlation functions take the following forms
\begin{eqnarray}
&&M^\mu(i\omega_n,\bQ) = \chi_{0}^{\mu,-1}(i\omega_n,\bQ) - \bar{I}_{\bQ} + \chi^{\mu,-1}_{loc}(i\omega_n,\bQ) \nonumber\\
&&\Sigma(i\omega_b,\bK) = G_{0}^{-1}(i\omega_n,\bK) - G^{-1}_{loc}(i\omega_n,\bK)
\end{eqnarray}
where the local correlation functions are defined as 
\begin{eqnarray}
&&\chi_{loc}^\mu (\tau,\bQ) = \langle T_\tau :S^\mu_{\bQ}:(\tau) :S^\mu_{-\bQ}:(0)\rangle_{S_{imp}} \nonumber \\ 
&&G_{loc}(\tau,\bK) = -\langle T_\tau c_{\bK,\sigma}(\tau) c_{\bK,\sigma}^\dag(0) \rangle_{S_{imp}} \, . 
\end{eqnarray} 
where we introduce the normal ordering of spin fields $:S^\mu_{\bQ}: = :S^\mu_{\bQ}: - m^\mu_{\bQ}$ with the magnetization $m^\mu_{\bQ}= \langle S^\mu_{\bQ}\rangle_{S_{imp}}$.  The bath functions and the effective magnetic fields are then defined self-consistently by requiring 
\begin{eqnarray}
&&\chi_{loc}^\mu (i\omega_n,\bQ) =\bigg(\frac{L_c}{L}\bigg)^d \sum_{\btq} \frac{1}{I_{\bQ+\btq} + M^\mu(i\omega_n,\bQ)} \nonumber\\
&&G_{loc}(i\omega_n,\bK) =\bigg(\frac{L_c}{L}\bigg)^d \sum_{\btk} 
\frac{1}{i\omega_n - \epsilon_{\bK+\btk} -\Sigma_{c}(i\omega_n,\bK)}
\nonumber \\ 
&& h_{\bQ}^\mu = - m^\mu_{-\bQ} 
\bigg( 
\chi_0^{\mu,-1}(i\omega_n=0,\bQ) + I_{\bQ}^\mu  - \bar{I}_{\bQ}^\mu 
\bigg) 
\label{eq:cedmft_eq}  
\end{eqnarray}
We note that, by adding the $h_{\bQ}^\mu$ contributions, we have normal-ordered the retarded interactions induced by the bosonic bath $\chi_0^{\mu}(\tau,\bQ)$.

The original lattice model are then solved by repeatedly solving the cluster model in Eq.~\ref{eq:KLimp} 
via quantum Monte Carlo until the self-consistent relations in Eq.~\ref{eq:cedmft_eq} are satisfied. Comparing to the single-impurity formula, we are able to partially capture the momentum dependency of the self energies and describe the inter-site correlations in CEDMFT. The inter-site correlations would allow attractive interactions between two electrons from different sites and, thus, provide a route towards unconventional superconductivity. Therefore, CEDMFT provides a platform to study the unconventional superconductivity induced by electron correlations. This approach has allowed the study of how Kondo destruction quantum criticality drives 
unconventional and high-$T_c$ superconductivity \cite{Hu2021.sc-x}.

\subsection{Models with disordered exchange interactions}

The EDMFT approach has some similarities with the dynamical
mean field theory of random spin systems.
The quantum Sherrington-Kirkpatrick model was considered by 
Bray and Moore in Ref.\,\cite{Bray80}, where the spin-exchange 
interactions are infinite-ranged, random and have a Gaussian distribution
whose mean vanishes and whose variance is 
$J^2/N_s$
where $N_s$ is the size of the system.
A disorder averaging using the replica trick,
as usual,
gives rise to 
 four-spin interactions, with each spin carrying a replica index.
Because the exchange couplings 
on different bonds are not correlated,
 the four-spin interaction 
 corresponds to two spins
at two different times $\tau$ and $\tau'$, from any site,
interacting with two other spins, 
at the same times $\tau$ and $\tau'$ and from
every other site. Taking the limit of $N_s \rightarrow \infty$
yields a single-site problem with retarded spin-spin
interactions. In the paramagnetic phase, the local correlators are determined 
by an effective local problem whose form is similar to
the spin part of 
Eq.\,(\ref{S-imp-kondo-lattice}), 
with a self-consistency
equation 
that corresponds to a semi-circular $\rho_{\rm I} (x)$.
When the spin symmetry is generalized from SU(2) to SU(N) and
the limit of large-$N$ is taken, a spin liquid ground state is realized \cite{SachYe93}.

More recently, a $t-J$ model with such a random-Gaussian $J$ interactions between the spins of the fermions has been studied. The saddle-point equations are similar to the EDMFT equations as reviewed in this article, with a semi-circular $\rho_I(x)$. In this case, while the asymptotic low-temperature dynamics will ultimately be similar to what happens in an SDW QCP, at intermediate temperatures and energies the results
\cite{Joshi2020,Dumitrescu2022,ChowdhuryRMP}
are similar to what we have surveyed.

\subsection{Application to electronic structure calculations: the GW$+$EDMFT approach }

The EDMFT approach has been combined with {\it ab initio} methods for electronic-structure calculations, incorporating non-perturbative effects in the GW-method (GW+DMFT). 
In the GW approach \cite{gw_origin}, 
the self-energies are derived by calculating the one-loop diagrams involving the screened Coulomb interaction, and are momentum- and frequency-dependent. 
The EDMFT method provides a non-perturbative approach for the local self-energies.
In the GW+EDMFT approach \cite{PhysRevLett.92.196402,PhysRevMaterials.1.043803,PhysRevLett.90.086402,kutepov2021spatial}, the two methods are combined where the local part of the GW self-energies are replaced by the EDMFT-derived local self-energies.
Thus, the GW+EDMFT approach provides a non-perturbative 
and efficient method to perform electronic structure calculations from first principles.

\subsection{EDMFT approach in new settings}

The EDMFT method is uniquely suited to elucidate 
the correlation physics in a variety of new settings. 
Here we describe two issues as examples.

One issue concerns  the
orbital-selective Mott phase, in which certain orbitals are
Mott localized while the others remain itinerant, which
is of considerable current interest in the field of iron-based superconductors \cite{Si16.1}.
Such an orbital-selective Mott phase 
in the presence of 
kinetic inter-orbital hybridization
has been demonstrated using certain auxiliary spin method
but has yet to be shown in methods based on the dynamical mean field theory. Recent renormalization-group analysis \cite{Hu-chen-dehyb22.x}
motivates an EDMFT-based examination of 
the interplay between the inter-orbital hybridization and 
intersite spin-exchange interactions, both of which are 
important to the physics of the iron-based superconductors.

Another issue deals with frustrated bulk systems with 
topological flat bands that are coupled to wide bands. 
Recent work has illustrated how an effective orbital-selective Mott transition can take place when an Anderson/Kondo lattice
model is realized as an effective description \cite{Hu-flat22.x}.
The EDMFT approach is expected to shed considerable new light on this emerging topic.


\section{Summary and Outlook}

We have provided an overview of the
 extended dynamical mean field theory for correlated electron
 models. The method is especially suited for settings where 
 the competing interactions arise from local and intersite terms, 
 respectively. Our emphasis has been on models for which the method was initially formulated and developed,
 namely models of the multi-band Anderson-lattice type 
 and those of 
 the
one-band Hubbard-Heisenberg kind. 

A setting in which this method has been most extensively used 
is the Kondo/Anderson lattice models for heavy fermion systems,
where the key issue is the 
dynamical competition between the local Kondo and 
intersite
RKKY interactions.
In elucidating how the amplitude of the Kondo singlet in the ground state and the weight of heavy quasiparticles are critically destroyed, these studies underline the notion
of Kondo destruction. The associated physics includes a loss 
of quasiparticles and dynamical Planckian scaling 
at the quantum critical point and a large-to-small Fermi surface jump across the quantum critical point~\cite{Hu-QCM-review2022}.

We have also discussed the prospect for further 
developing the 
method, including in the form that is suitable to study
unconventional superconductivity. In the same spirit,
we note that new settings have continue to arise for which
the approach of the extended dynamical mean field theory 
is poised to clarify new physics. We have described two such 
issues. One concerns the 
orbital-selective Mott phase in the iron-based superconductors,
and the other about the correlation physics of coupled topological
flat-wide bands in kinetically-frustrated bulk systems.
Given the versatility that the EDMFT approach has shown in 
elucidating the dynamical interplay between local and spatial
correlations, 
we expect that other settings will also develop in which the
approach will provide new insights into the underlying correlation physics.


\acknowledgements

We would like to thank the late D. R. Grempel, 
A. Cai, M. T. Glossop, P. Goswami,
 A. Kandala, K. Ingersent, S. Kirchner, E. M. Nica, J. H. Pixley,  J. L Smith, J.-X. Zhu, L. J. Zhu and, particularly,
 S. Paschen and F. Steglich,
 for collaborations on work pertaining
 to the subject of this article.
 This work has been supported in part by
the NSF Grant No.\ DMR-2220603 (Q.S.)
the Air Force Office of Scientific Research under 
Grant No.\ FA9550-21-1-0356 (H.H.),
the Robert A. Welch Foundation Grant No.\ C-1411 (L.C.), and the 
European Research Council (ERC) under the European Union’s Horizon 2020 research and innovation program (Grant Agreement No. 101020833) (H.H).


\newpage
\bibliographystyle{naturemagallauthors}
\bibliography{EDMFT_review}

\end{document}